\documentclass[a4paper,11pt]{article}

\usepackage{jheppub} 

\usepackage[T1]{fontenc} 
\usepackage{amsmath}
\usepackage{epstopdf}

\title{Novel black holes in higher derivative gravity} 

\author[a]{Yang Huang}
\author[b]{Dao-Jun Liu}
\author[a,1]{Hongsheng Zhang,\note{Corresponding author.}}

\affiliation[a]{School of Physics and Technology, University of Jinan, 336, West Road of Nan Xinzhuang, Jinan 250022, Shandong, China}
\affiliation[b]{Department of Physics, Shanghai Normal University, 100 Guilin Road, Shanghai 200234, China}

\emailAdd{sps\_huangy@ujn.edu.cn}
\emailAdd{djliu@shnu.edu.cn}
\emailAdd{sps\_zhanghs@ujn.edu.cn}

\abstract{We find a class of novel black holes in higher derivative theory. The novel black holes follow behavior of~\sch\ ones at large mass limit, while dramatically differentiate from ~\sch\ ones for little holes because of the effects which may root in quantum gravity. The temperature of the hole takes  maximum for a specific mass, which is related to the new sale introduced in the higher derivative theory, and goes to zero at little mass limit. This property leads to a significant observation that the novel black hole may be a candidate for dark matters evading constraint from $\gamma$-ray burst.}

\usepackage{hyperref}
\hypersetup
{
	colorlinks,%
	citecolor=blue,%
	linkcolor=magenta,%
	urlcolor=blue,%
}

\begin{document}
	\newcommand{\be}{\begin{equation}}
	\newcommand{\en}{\end{equation}}
	\newcommand{\bea}{\begin{eqnarray}}
	\newcommand{\ena}{\end{eqnarray}}
	\newcommand{\sch}{Schwarzschild}
	\newcommand{\wdt}{\widetilde}
	\newcommand{\red}[1]{\textcolor{red}{#1}}
	\newcommand{\green}[1]{\textcolor{green}{#1}}
	\newcommand{\blue}[1]{\textcolor{blue}{#1}}
	\newcommand{\cyan}[1]{\textcolor{cyan}{#1}}
	\newcommand{\purple}[1]{\textcolor{purple}{#1}}
	\newcommand{\yellowbox}[1]{\colorbox{yellow}{#1}}
	\newcommand{\purplebox}[1]{\colorbox{purple}{#1}}
	\newcommand{\yellow}[1]{\textcolor{yellow!70!red}{#1}}
	\title{Novel black holes in higher derivative gravity}

	\maketitle
	
	
	\section{Introduction}
	There are several motivations to consider gravity theories with higher derivative terms.
	Remarkably, the renormalizability of gravity requires higher derivative terms, especially the quadratic curvature terms \cite{PhysRevD.16.953}. On the other side, the quadratic curvature term appears naturally when one considers the renormalization of stress energy \cite{TSBunch_1979,PhysRevD.17.1477,Wald:1977up,Birrell:1982ix}. Some unification theories also lead to  quadratic curvature terms in low energy effective gravity theory \cite{PhysRevD.36.392,polchinski_1998}. The causal issue of quadratic gravity is discussed in \cite{Edelstein:2021jyu}.
	
	In principle there may exist spherical vacuum solutions other than \sch~ in modified gravities since vacuum quadratic gravity equals to Einstein gravity with some form of stress energy and thus  Birkhoff theorem is invalid. Such solutions have been found, for examples, in \cite{Zhang:2021cea,Zhang:2021sjx,Zhang:2014ala,Zhang:2014kla}.
	
	The first non-\sch\ black hole solution for vacuum quadratic gravity was obtained numerically in \cite{Lu:2015cqa}. Then, another sector of this non-\sch\ solution was obtained in \cite{PhysRevD.97.024015}. Actually, the solutions in Refs.\cite{Lu:2015cqa,PhysRevD.97.024015} belong to the same branch of the non-\sch\ black holes, since their curves in the parameter space intersect with the \sch\ solution at the same point $r_0/l_\alpha\approx0.876$, where $r_0$ is the event horizon of the black hole, and $l_\alpha$ is a length scale. Stability of the non-\sch\ black hole was studied in \cite{Cai:2015fia,Held:2022abx}. Other spherical solutions in quadratic gravity were studied in \cite{Novikov1965,Podolsky:2019gro,Pravda:2020zno,Svarc:2018coe}. Charged black holes in the Einstein-Maxwell-Weyl gravity and the quasinormal modes were obtained in \cite{Lin:2016jjl,Wu:2019uvq,Zou:2020msk}. Approximate analytical solutions in quadratic gravity is discussed in \cite{PhysRevD.102.124026,Sajadi:2022tgi}. In principle, the quadratic term implies high energy behavior of the theory, and the theory reduces to general relativity (Einstein-Hilbert term) at low energy regime. For the non-\sch\ solutions in \cite{Lu:2015cqa,PhysRevD.97.024015}, the black hole mass decreases with the increase of the horizon radius, and eventually the mass reaches a negative value. This behaviour has nothing similar to a \sch~one, while \sch~black hole should be reproduced for large holes, i.e., the low energy limit. Thus it is a pivotal task to explore a solution matching the theoretical expectation of quadratic gravity. In the followings, a novel non-\sch\ black holes are obtained. We find the novel black holes perfectly satisfy the theoretical expectation. Furthermore, the temperature of the novel black hole goes to zero rather than be divergent at low mass limit. Theoretically, the remnant is able to eternally hide information of its progenitor. In observational aspect, it can serve as dark matter composed by small mass primordial black holes which naturally evades stringent constraint from observation of $\gamma$-ray burst.
	
	\section{Field equations}
	We start from the most general action in four dimensional Einstein-Hilbert gravity with added quadratic curvature terms $I=\int d^4x\sqrt{-g}\left(\gamma R-\alpha C_{abcd}C^{abcd}+\beta R^2\right)$, where $\alpha,\ \beta$ and $\gamma$ are constants and $C_{abcd}$ is the Weyl tensor \cite{Lu:2015cqa}. We choose $\gamma=1$, and the equations of motion are
	\begin{equation}\label{Eq: EOM}
	\begin{aligned}
	R_{ab}&-\frac{1}{2}g_{ab}R-4\alpha B_{ab}+2\beta\left(R_{ab}-\frac{1}{4}g_{ab}R\right)\\&+2\beta\left(g_{ab}\Box R-\nabla_a\nabla_bR\right)=0,
	\end{aligned}
	\end{equation}
	where $B_{ab}=\left(\nabla^c\nabla^d+\frac{1}{2}R^{cd}\right)C_{acbd}$ is the Bach tensor. For static black holes, the Ricci scalar must vanish, and the field equation can be greatly simplified by setting $\beta=0$ without loss of generality \cite{PhysRevD.82.104026}. Parameter $\alpha$ has the dimension of length square, and can be written as $\alpha=l^2_\alpha/2$, where $l_\alpha$ is the length scale of the theory \cite{PhysRevD.97.024015}. All physical quantities are measured by $l_\alpha$. \sch\ black hole is a solution of the theory, since the vanishing Ricci scalar, together with $R_{ab}=0$, obeys the field equation (\ref{Eq: EOM}). It was suggested in \cite{Lu:2015cqa} that there might exist static, spherically symmetric non-\sch\ black hole solutions with non-vanishing Ricci tensor. Nevertheless, finding black hole solution is non-trivial as the forth order equations of motion are numerically unstable. We introduce a new method to investigate the numerical solutions in quadratic gravity.
	
	\section{Numerical methods}
	Consider static and asymptotically flat black holes, the metric can be written as
	\begin{equation}\label{Eq: metric 2}
	\begin{aligned}
	ds^2=&\frac{l^2_\alpha dr^2}{N(r)e^{2B}}+l^2_\alpha r^2\left(d\theta^2+\sin^2\theta d\phi^2\right)\\
	&-N(r)e^{2A}dt^2,\ \text{with}\ N(r)=1-\frac{r_0}{r},
	\end{aligned}
	\end{equation}
	where $\left\lbrace A,B\right\rbrace $ are functions of $r$, and $r_0$ is the event horizon. It should be noted that both $r$ and $r_0$ are dimensionless. With this ansatz, the field equation (\ref{Eq: EOM}) yields two coupled ordinary differential equations of $\left\lbrace A,B\right\rbrace$ (see Appendix.\ref{Appedix: eqsAB} for more details). The \sch\ solution is described by $A=B=0$ for any values of $r_0$, while for non-\sch\ black holes, functions $\left\lbrace A,B\right\rbrace $ are non-vanishing except at infinity.
	
	For the methods in Refs.\cite{Lu:2015cqa,PhysRevD.97.024015}, one can only integrate the field equations from $r_0$ to some point not far from the event horizon before the metric functions diverge. To this end, it is useful to introduce a compactified coordinate
	\begin{equation}
	x=\frac{r-r_0}{r+L},
	\end{equation}
	where $L$ is an free positive parameter. The final physical solutions are independent on the choice of $L$. The compactified coordinate maps $r\in[r_0,+\infty)$ to a finite region $[0,1]$, where the event horizon is located at $x=0$, and spatial infinity is at $x=1$. The compactified coordinate has at least three advantages: (i) it allows one to include the infinity into the computational domain; (ii) the imposition of boundary conditions at infinity becomes more simple and precise; (iii) the interpolation of the numerical solutions at infinity is of high accurate, which improves the accuracy of the physical quantities extracted at infinity.
	
	The boundary conditions of the problem are as follows. First, for asymptotically flat solutions, both $g_{tt}$ and $g_{rr}$ tend to $1$ at infinity. Thus, the outer boundary conditions are $A(1)=B(1)=0$. Second, near the event horizon, a power series expansion of $A$ and $B$ can be obtained of the form, $A=a_0+a_1 x+\mathcal{O}\left(x^2\right)$ and $B=b_0+b_1 x+\mathcal{O}\left(x^2\right)$, where $a_0$ is a trivial parameter whose value can be used to rescale the time coordinate. Then, substituting the expansions into the equations of $A$ and $B$, we have
	\begin{equation}
	\begin{aligned}
	&\left[A'+\left(r_0+L\right)e^{-B}\sinh(B)\left(\frac{1}{r_0}+\frac{r_0}{4}e^{-2B}\right)\right]\Bigg|_{x=0}=0,\\
	&\left[B'+\left(r_0+L\right)e^{-B}\sinh(B)\left(\frac{1}{r_0}-\frac{3r_0}{4}e^{-2B}\right)\right]\Bigg|_{x=0}=0,
	\end{aligned}
	\end{equation}
	where a prime denotes the derivative with respect to $x$.
	
	The non-\sch\ black holes have two non-trivial parameters, $r_0$ and $\delta=e^{2b_0}-1$, where $\delta$ describes the deviation from \sch. The numerical black hole solutions are constructed by the spectral method together with the Newton-Raphson method. In order to check the validity of our code, we have recovered all the non-\sch\ solutions in \cite{PhysRevD.97.024015,Lu:2015cqa}. Moreover, a novel branch of non-\sch\ black holes with some interesting properties are obtained via the same code. For convenience, we use \textit{solutions I (II)} to refer to the non-\sch\ black hole in Refs.\cite{Lu:2015cqa,PhysRevD.97.024015} (this work) in the following text.
	
\begin{figure*}
		\centering	
		\includegraphics[width=0.45\textwidth,height=0.36\textwidth]{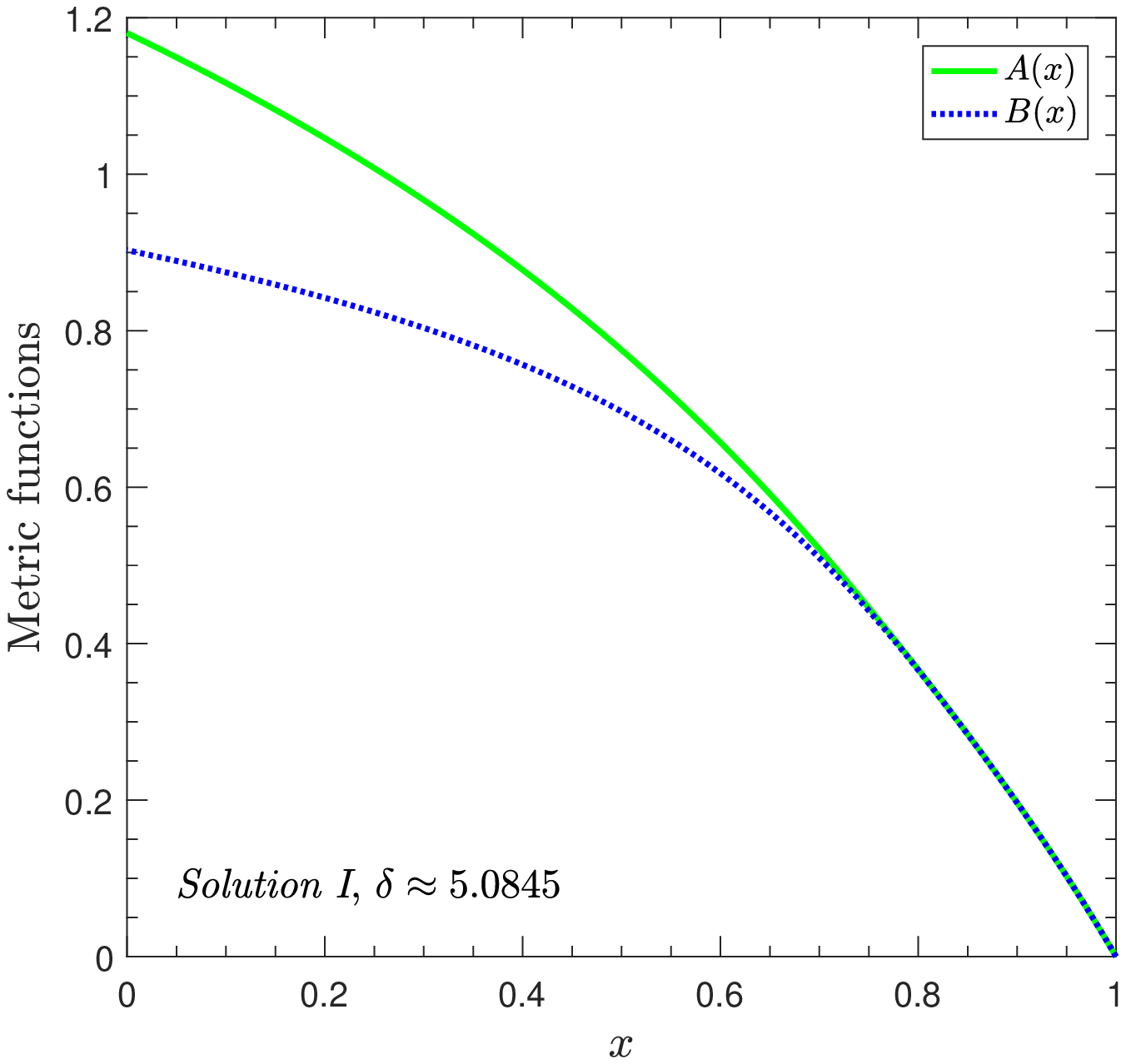}	
		\includegraphics[width=0.45\textwidth,height=0.36\textwidth]{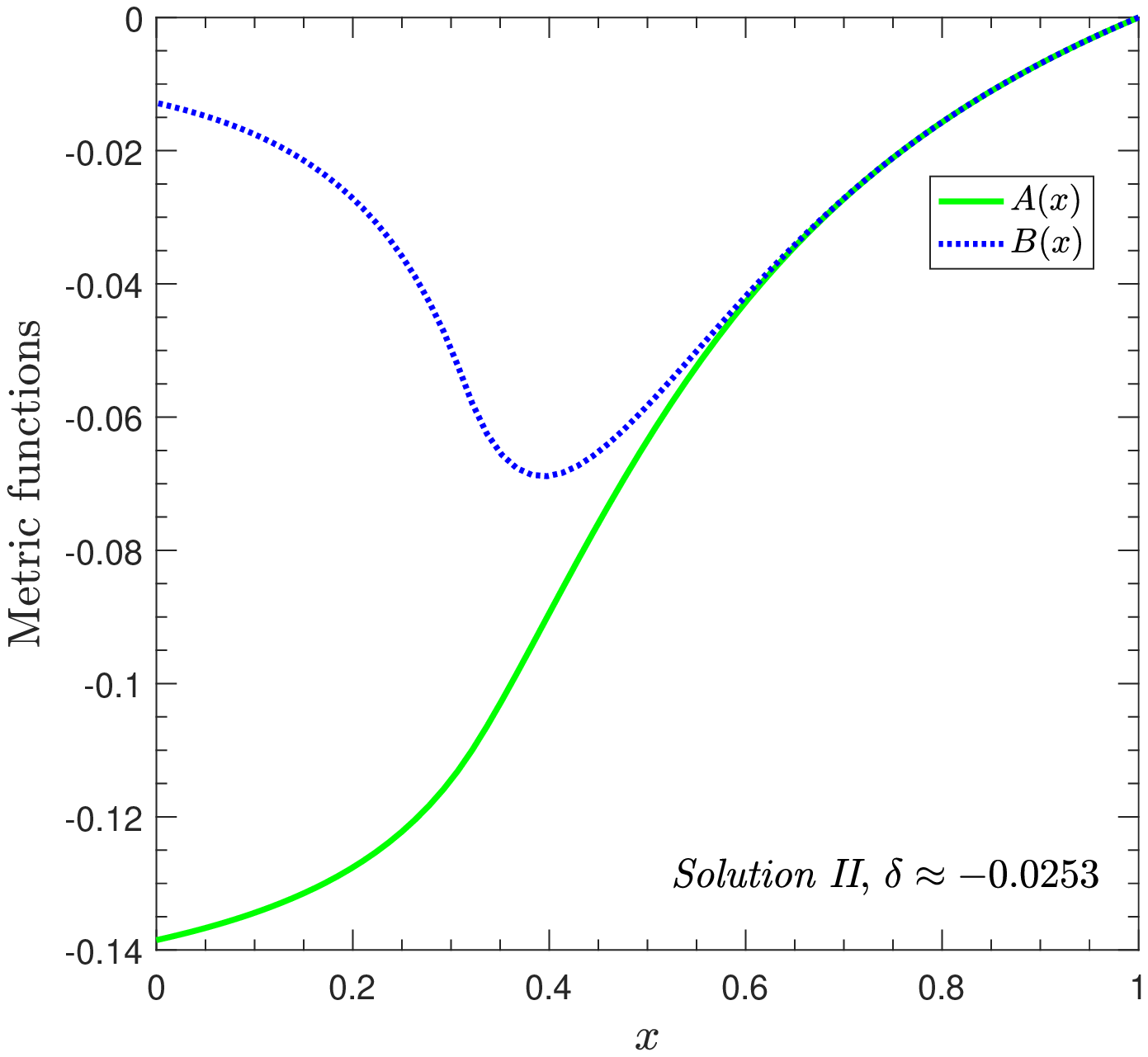}	
		
		\includegraphics[width=0.45\textwidth,height=0.36\textwidth]{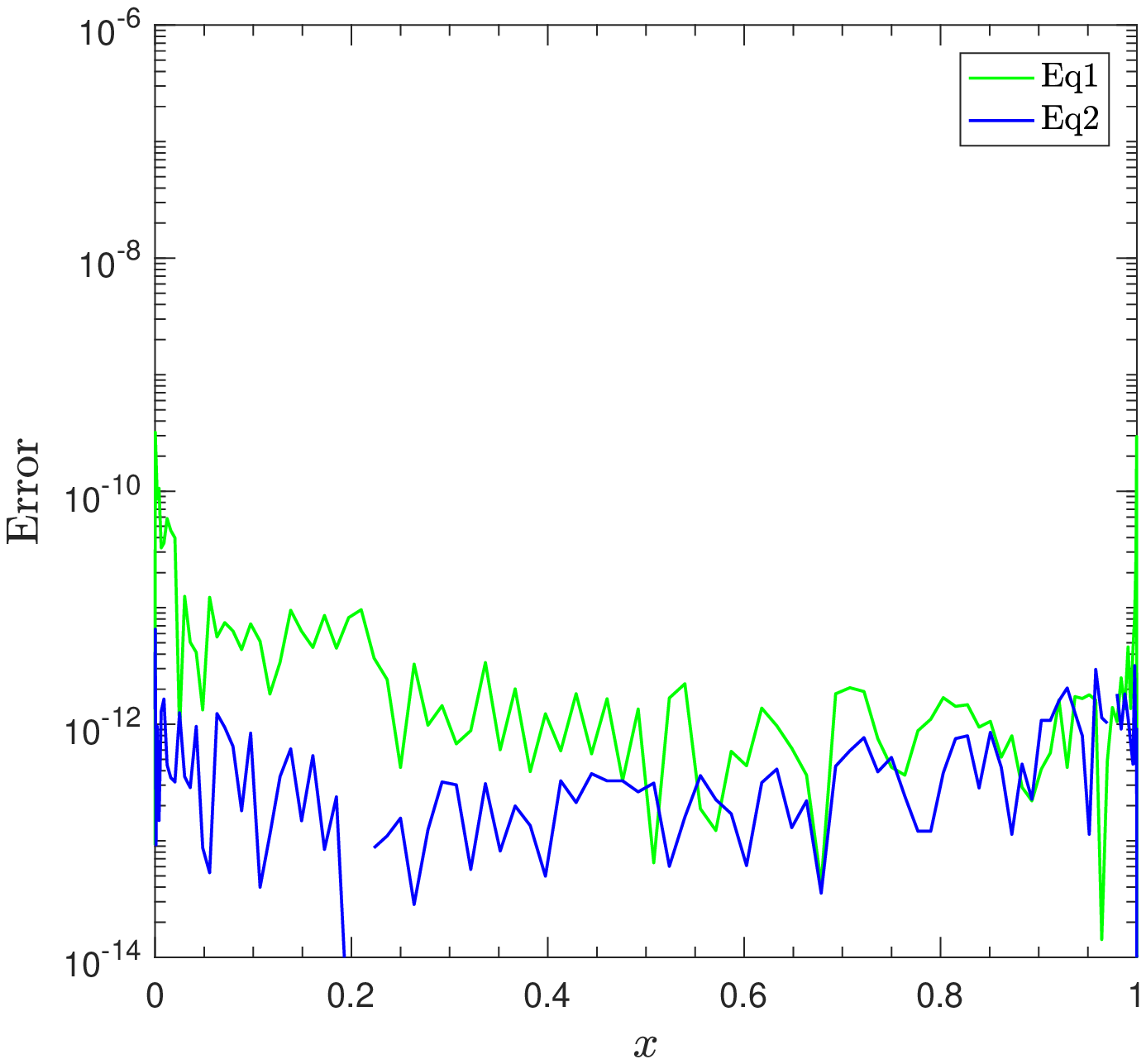}
		\includegraphics[width=0.45\textwidth,height=0.36\textwidth]{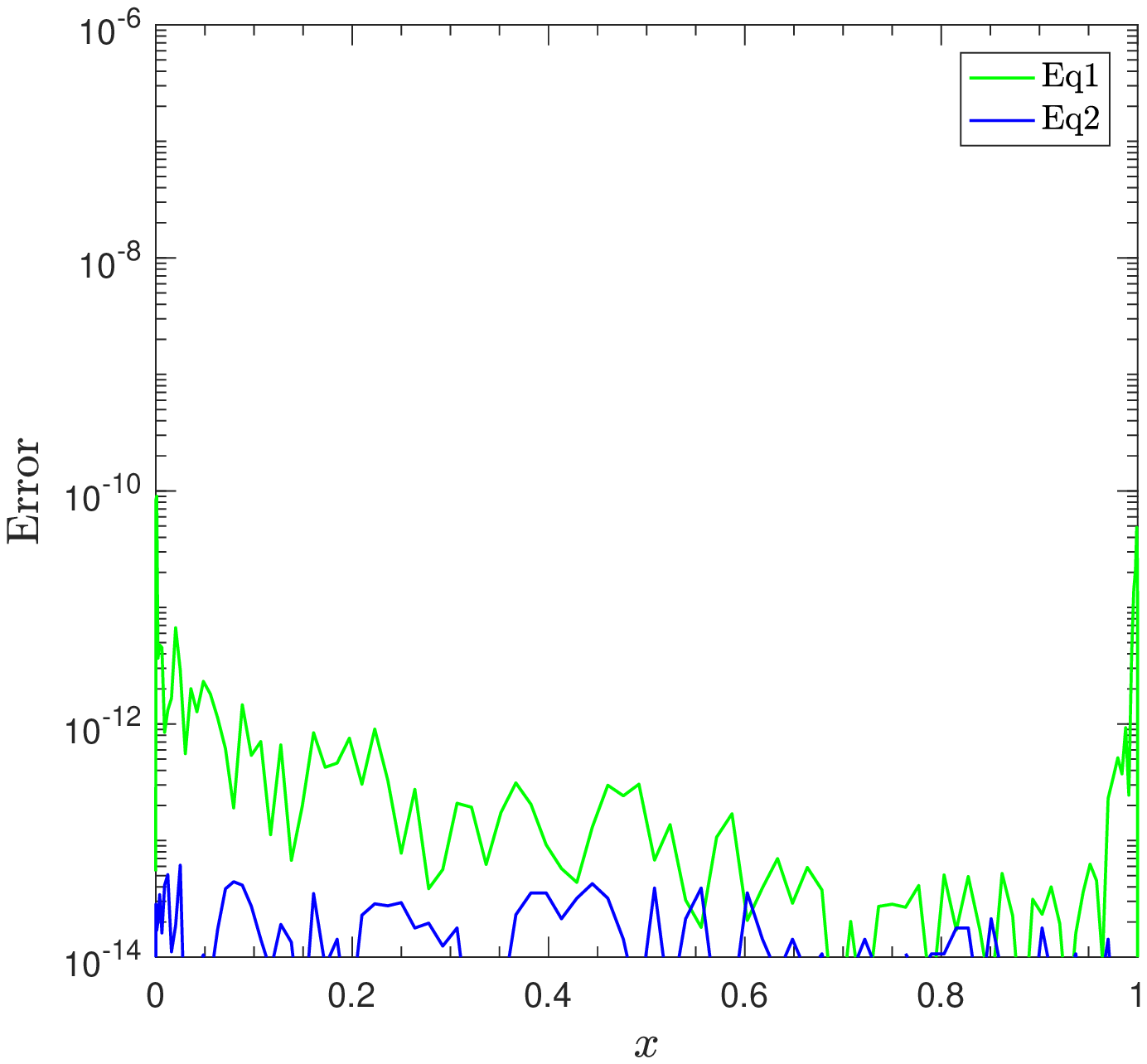}
		\caption{Two non-\sch\ black holes for $r_0=2$. Top panel shows numerical results of functions $\left\lbrace A,B\right\rbrace$, and bottom panel shows the absolute value of the error in Eqs.(\ref{Eq: Eq1}) and (\ref{Eq: Eq2}) (see Appendix.\ref{Appedix: eqsAB}). The left and right panels correspond to \textit{solutions I and II}, respectively.}
		\label{Fig: sol1 vs sol2}
	\end{figure*}

\begin{figure*}
		\centering	
		\includegraphics[width=0.5\textwidth,height=0.43\textwidth]{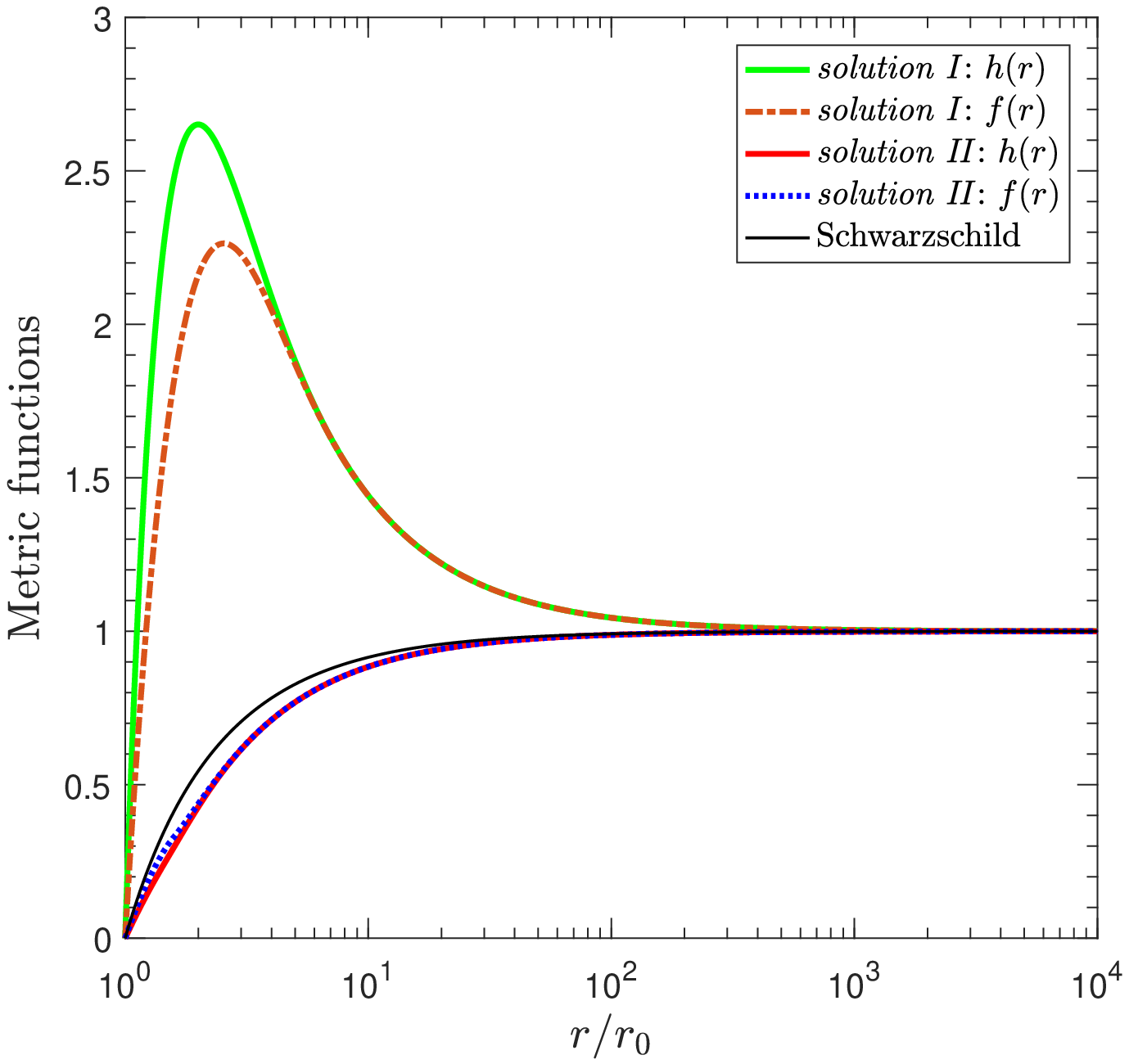}	
		\caption{Functions $\left\lbrace h,f \right\rbrace $ of the two non-\sch\ black holes with $r_0=2$. For \sch\ black hole, $h=f=1-r_0/r$ (black solid line).}
		\label{Fig: f vs h}
	\end{figure*}

	Figure \ref{Fig: sol1 vs sol2} shows the two non-\sch\ black holes for $r_0=2$. We see that numerical errors of the field equations are below $\mathcal{O}\left(10^{-10}\right)$. Clearly, the relations of metric functions $\left\lbrace h,f\right\rbrace $ in Refs.\cite{Lu:2015cqa,PhysRevD.97.024015} and $\left\lbrace A,B\right\rbrace$ are given by
	\begin{equation}
	h(r)=N(r)e^{2A(r)},\;\;\;f(r)=N(r)e^{2B(r)}.
	\end{equation}
	In Fig.\ref{Fig: f vs h}, we exhibit $\left\lbrace h,f\right\rbrace $ of the two non-\sch\ black holes as functions of the radial coordinate $r$. One may check that lines of \textit{solution I} are consistent with the right panel of Fig.2 in Ref.\cite{Lu:2015cqa}. The difference is that our numerical solutions can be interpolated to spatial infinity, and the plot shows clearly that $\lim\limits_{r\rightarrow\infty}h(r)=\lim\limits_{r\rightarrow\infty}f(r)=1$. Therefore, our results provide a firmer evidence for the existence of asymptotically flat non-\sch\ black holes. More significantly, we show that for a given $r_0$, there exist at least two non-\sch\ black hole solutions.
	
\begin{figure}
		\centering	
		\includegraphics[width=0.45\textwidth,height=0.36\textwidth]{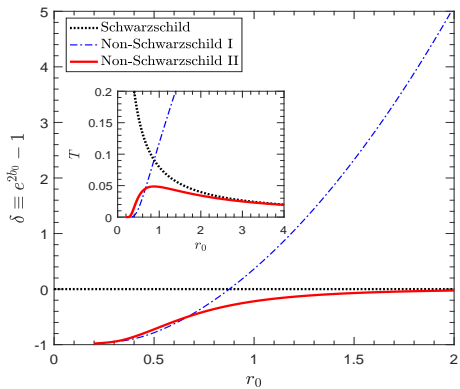}	
		\caption{$\delta$ vs $r_0$ for (non-)\sch\ black holes. The blue curve correspond to the branch of non-\sch\ black holes in Refs.\cite{Lu:2015cqa,PhysRevD.97.024015}, which intersect with the \sch\ solution ($\delta=0$, black dotted) at $r_0\approx 0.876$. The red solid curve is the non-\sch\ solution in this work. (Inset) The Hawking temperature $T$ vs $r_0$ around intersection points.}
		\label{Fig: delta vs r0}
	\end{figure}

	We have obtained two non-\sch\ black holes for $r_0=2$. Starting from them, we could find two branches of non-\sch\ black holes by sweeping the $r_0$ space. In Fig. \ref{Fig: delta vs r0}, we plot the parameter space of \textit{solutions I and II}. Both \textit{solutions I and II} have their own lower bound for parameter $r_0$. As usual, the Hawking temperature is given by surface gravity at the horizon $e^{A(r_0)+B(r_0)}/(4\pi r_0)$. Approaching the minimum value of $r_0$, the Hawking temperature tends to zero (see the inset) and the numerics becomes challenging. This situation is similar to the case of extreme black holes in general relativity. Note that for \textit{solutions I}, the Hawking temperature monotonously increases with $r_0$, which implies that a larger black hole has a higher temperature. This hole  cannot come back to a \sch~one for large radius. This result is rather puzzling when one examines the action, which implies the theory reduces to general relativity (Einstein-Hilbert action) at low energy limit, or say large black hole. Focusing on \textit{solutions II} (red line in Fig.\ref{Fig: delta vs r0}), we see that $\delta$ is always negative and the temperature is lower than \sch\ for the same values of $r_0$. When $r_0$  increases, as $\delta$ monotonously approaches to zero, the black hole gradually reduces to \sch. This result completely satisfies the normal property of the action.
	
	Having established the numerical solutions of non-\sch\ black holes, we explore their physical properties. Quantities of interest are defined by the metric functions. For asymptotically-flat black holes, the  Arnowitt-Deser-Misner (ADM) mass $M$ can be read off from the asymptotic behavior: $g_{tt}\simeq-1+2M/r$. One can show that the Wald entropy for the metric (2) in quadratic gravity is $S=\pi r_0^2-2\pi\delta$ \cite{Fan:2014ala}.
	
\begin{figure}
		\centering		
		\includegraphics[width=0.45\textwidth,height=0.36\textwidth]{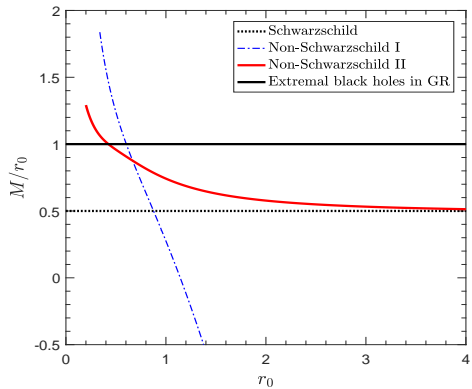}	
		\caption{Mass-to-size ratio $M/r_0$ of the two branches of non-\sch\ black holes as functions of $r_0$. Curves are marked in the same style as in Fig.\ref{Fig: delta vs r0}.}
		\label{Fig: masstosize vs r0}
	\end{figure}

	Figure \ref{Fig: masstosize vs r0} shows the mass-to-size ratio $M/r_0$ of the non-\sch\ black holes as functions of $r_0$. For small $r_0$, both \textit{solutions I and II} violate the extreme bound in general relativity. As is known, black holes in Kerr-Newman family obey $M/r_0\leq1$. With the increase of $r_0$, the ratio $M/r_0$ of both \textit{solutions I and II} decreases monotonously. For \textit{solutions I}, the ratio $M/r_0$ becomes negative for $r_0>r^{m=0}_0\approx1.143$. For \textit{solutions II}, the ratio $M/r_0$ goes to $1/2$ at large $r_0$, which is consistent with the results in Fig.\ref{Fig: delta vs r0} that \textit{solutions II} reduce to \sch\ in the regime of large $r_0$.
	
\begin{figure*}
		\centering	
		\includegraphics[width=0.45\textwidth,height=0.36\textwidth]{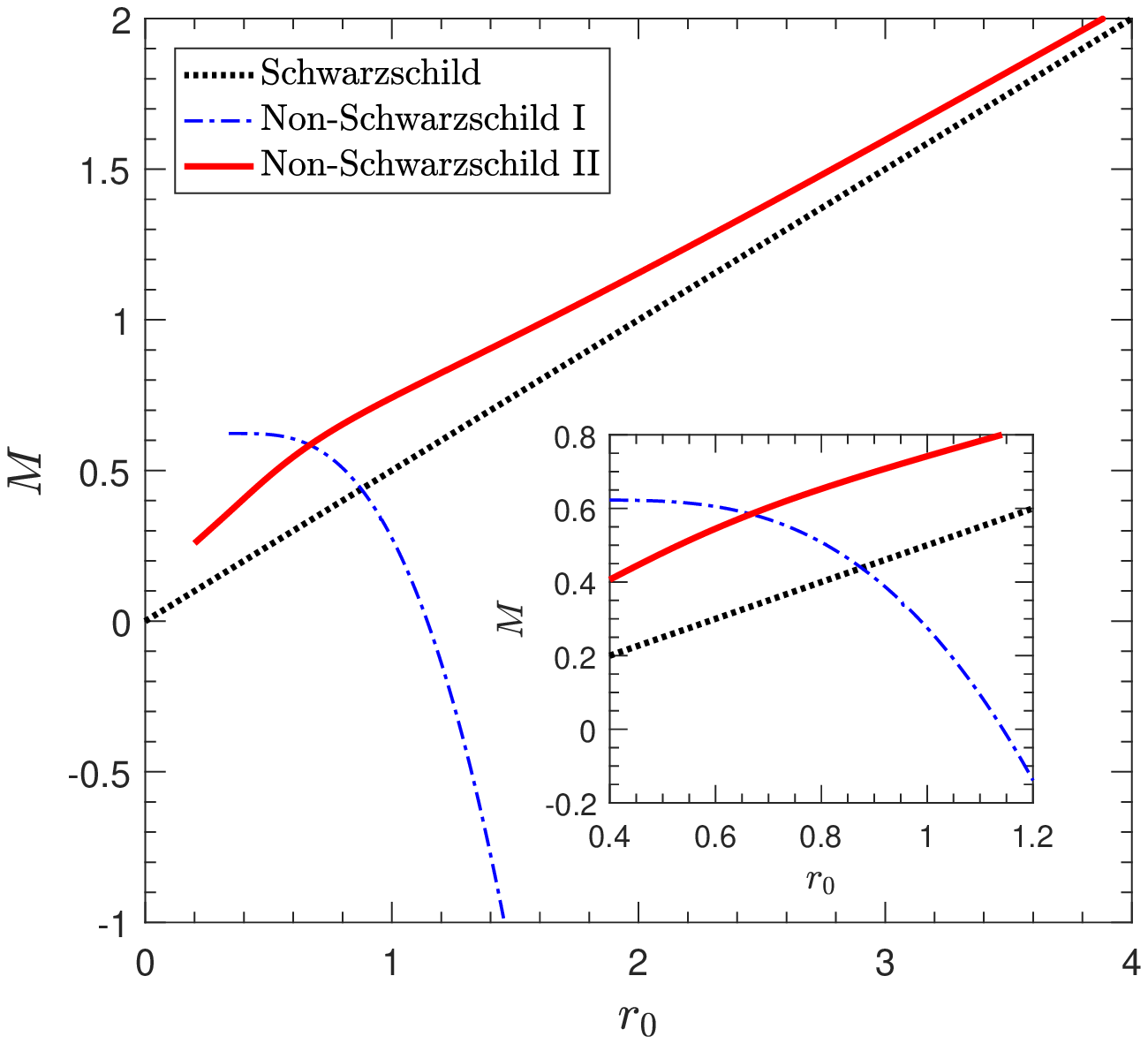}	
		\includegraphics[width=0.45\textwidth,height=0.36\textwidth]{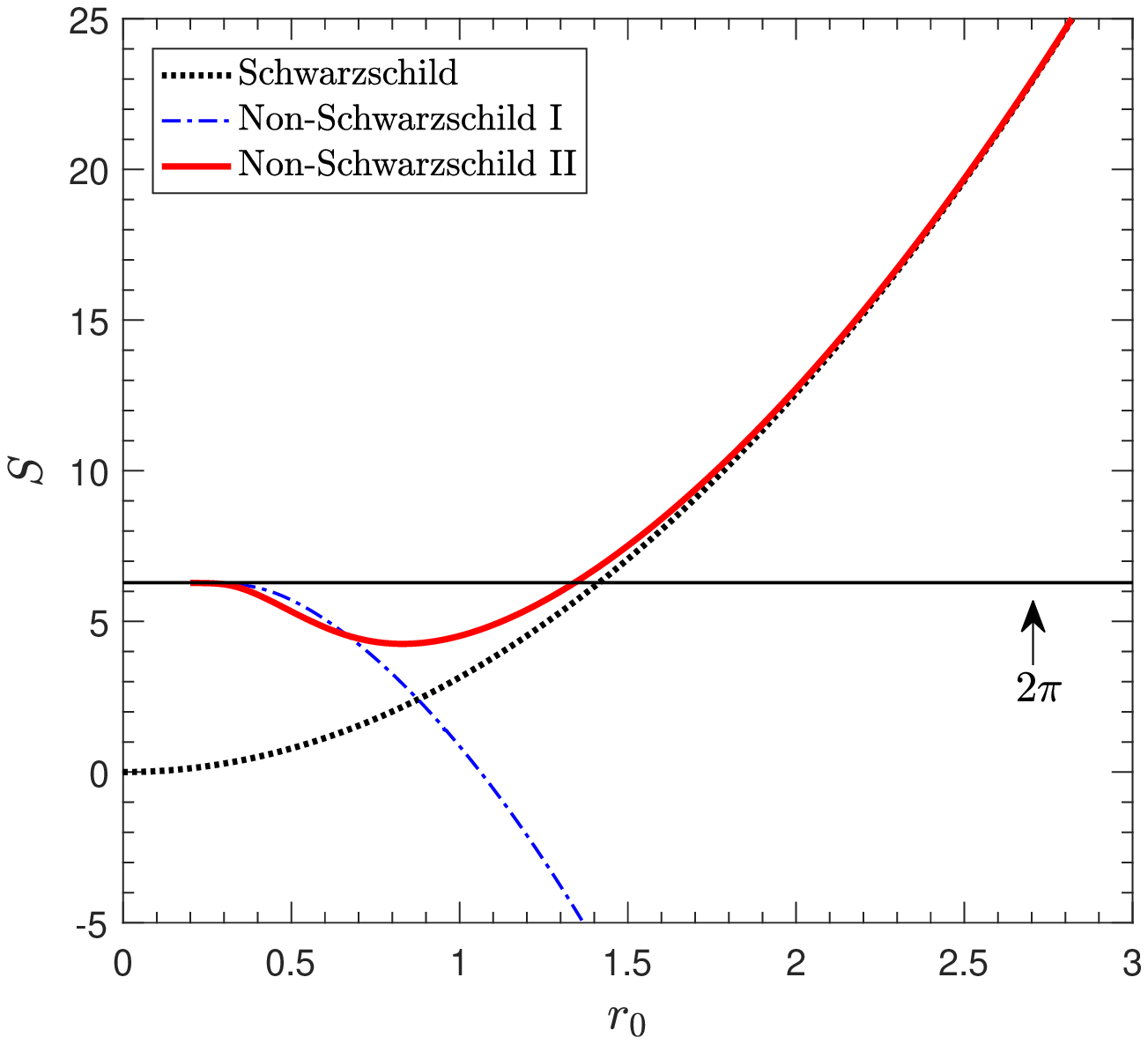}	
		\caption{The ADM mass (left panel) and entropy (right panel) as functions of $r_0$. Curves are marked in the same style as in Fig.\ref{Fig: delta vs r0}. The inset in left panel displays masses of different solutions around intersection points.}
		\label{Fig: mass vs r0}
	\end{figure*}

	In Fig.\ref{Fig: mass vs r0}, we exhibit the ADM mass and entropy of \textit{solutions I and II} as functions of $r_0$. When $r_0$ is small, both \textit{solutions I and II} have a finite mass $M$, and their entropy $S$ tend to $2\pi$. As $r_0$ increases, however, the two solutions behave differently. For \textit{solutions I}, both $M$ and $S$ decrease as $r_0$ increases, and will eventually be negative. This means that black hole with small horizon radius could have a larger ADM mass. For \textit{solutions II}, however, the mass and entropy are always positive, and $M$ monotonously increases with $r_0$.
	
\begin{figure}
		\centering	
		\includegraphics[width=0.45\textwidth,height=0.36\textwidth]{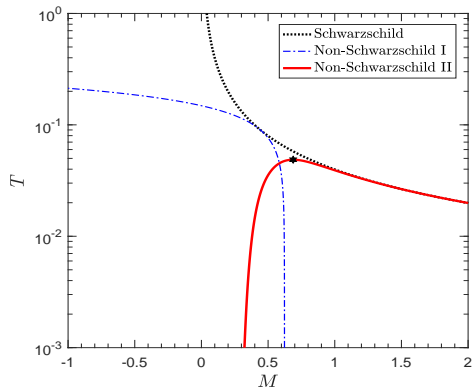}	
		\caption{Hawking temperature as functions of the ADM mass. Curves are marked in the same way as in Fig.\ref{Fig: delta vs r0}.}
		\label{Fig: M vs T}
	\end{figure}

	Figure \ref{Fig: M vs T} shows the Hawking temperature for \textit{solutions I and II} as functions of $M$. Both for \sch\ and \textit{solutions I}, the temperature is increased with the decrease of $M$, i.e., a more massive black hole may has a lower temperature. It is known that the \sch\ black hole is thermodynamically unstable. Focus on the dot-dashed blue line Fig.\ref{Fig: M vs T}, consider a black hole of \textit{solutions I} initially with a positive mass $M$ and a relatively low temperature. As the black hole evaporates, the mass decreases and the temperature gets higher. Eventually $T$ reaches a finite value, while the mass and entropy become negative (see Fig. \ref{Fig: mass vs r0}). By definition, the entropy of a system is the logarithm of the number of microstates. A negative entropy is unphysical, since it does not correspond to any microstates.
	
	For \textit{solutions II} (solid red line in Fig.\ref{Fig: M vs T}), there is a maximum temperature, $T_{\text{max}}\approx0.049l^{-1}_{\alpha}$. At right side of the maximum, the black hole described by \textit{solutions II} is thermodynamically unstable, just as a \sch~one. At left side of maximum, the black hole becomes thermodynamically stable. Its temperature becomes lower for a lighter black hole. This is a novel effect rooting in the quadratic term, which further may be yielded by quantum effects. The lifetime of the \textit{solutions II} black hole is eternal.
	
	\section{Conclusions}
	Hawking radiation carries problems as much as it solved \cite{Hawking:1974rv}. The most prominent problem is the unitary puzzle, or black hole information paradox. One of main approaches for black hole information paradox is remnant assumption, in which a residue Planck-mass object is assumed to carry all the information of the original black hole. There are several heuristic investigations of the remnant assumption \cite{AHARONOV198751,Giddings:1992hh,Banks:1992is,Xiang:2006mg,Maziashvili:2005pp,Chen:2012tx,Horowitz:2003he,Xiang:2013sza,Ali:2014xqa,Olmo:2013gqa}. Almost all previous studies concentrate on qualitative analysis of the remnant. A solid theoretical quantitative description of the remnant  is still in want. We present a quantitative solution in quadratic gravity. Quadratic term is required in quantum theory of gravity. Black hole paradox is also expected to be completely solved in a full-fledged quantum theory of gravity. Thus it is not surprising if quadratic gravity encodes critical information for black hole information paradox. Our \textit{solutions II} may pave a promising road to decode this relation.
	
	One more interesting point is that \textit{solution II} serves as a reasonable candidate of dark matter of primordial black hole. The permitted parameter spaces of hypothetical dark matter particles are compressed into a tiny corner. With detection of more and more gravitational wave events, the idea that primordial black holes play the role of dark matter is studied seriously. Primordial black hole is collapsed in the early universe because of cosmic perturbation or some phase transitions \cite{Liu:2019lul,Liu:2021svg,Hashino:2022tcs,Novikov1965,Hawking:1971ei,Carr:1974nx}. Different from astrophysical black hole, in principle there is no mass constraint for primordial black hole.  The abundance of primordial black holes in different mass intervals is constrained by astrophysical observations. The primordial black holes less than $10^{13}$kg have stringent constraint from observations of $\gamma$-ray burst, as Hawking radiations of primordial black holes \cite{PhysRevD.101.023010}. Clearly, black hole of \textit{solutions II} successfully evades the constraint from $\gamma$-ray burst if one set $l_{\alpha}>1.08\times10^{-14}$m, as the scale of a proton.
	
	We make a short summary of this work. We find a novel branch of non-\sch\ black holes (\textit{solutions II}) in quadratic gravity. \textit{Solutions II} reduces to \sch~in large mass limit since Einstein-Hilbert action is in domination, and shows some marvelous behavior for little black holes since quadratic term is in domination. The temperature of black hole of \textit{solutions II} goes to zero when its mass tends to zero. This interesting behavior may root in quantum effects of gravity, which requires quadratic term. \textit{Solutions II} black hole cannot be burned out through Hawking radiations. It is possible to  achieve two aims at one stroke by considering the remnant \textit{solutions II}. It may be a hopeful object accounting for black hole paradox, as well as for dark matter.
	
	%

	\begin{acknowledgments}
		Our thanks go to Prof. Hong L\"{u} for helpful discussions. This work is supported by the National Natural Science Foundation of China Grants Nos.12275106 and 12235019.
	\end{acknowledgments}

	\bibliographystyle{unsrt}
	\bibliography{bhsol}

\begin{thebibliography}{10}

\bibitem{PhysRevD.16.953}
K.~S. Stelle.
\newblock Renormalization of higher-derivative quantum gravity.
\newblock {\em Phys. Rev. D}, 16:953--969, Aug 1977.

\bibitem{TSBunch_1979}
T~S Bunch.
\newblock On renormalisation of the quantum stress tensor in curved space-time
  by dimensional regularisation.
\newblock {\em Journal of Physics A: Mathematical and General}, 12(4):517, apr
  1979.

\bibitem{PhysRevD.17.1477}
Robert~M. Wald.
\newblock Trace anomaly of a conformally invariant quantum field in curved
  spacetime.
\newblock {\em Phys. Rev. D}, 17:1477--1484, Mar 1978.

\bibitem{Wald:1977up}
Robert~M. Wald.
\newblock {The Back Reaction Effect in Particle Creation in Curved Space-Time}.
\newblock {\em Commun. Math. Phys.}, 54:1--19, 1977.

\bibitem{Birrell:1982ix}
N.~D. Birrell and P.~C.~W. Davies.
\newblock {\em {Quantum Fields in Curved Space}}.
\newblock Cambridge Monographs on Mathematical Physics. Cambridge Univ. Press,
  Cambridge, UK, 2 1984.

\bibitem{PhysRevD.36.392}
Robert~C. Myers.
\newblock Higher-derivative gravity, surface terms, and string theory.
\newblock {\em Phys. Rev. D}, 36:392--396, Jul 1987.

\bibitem{polchinski_1998}
Joseph Polchinski.
\newblock {\em String Theory}, volume~1 of {\em Cambridge Monographs on
  Mathematical Physics}.
\newblock Cambridge University Press, 1998.

\bibitem{Edelstein:2021jyu}
Jose~D. Edelstein, Rajes Ghosh, Alok Laddha, and Sudipta Sarkar.
\newblock {Causality constraints in Quadratic Gravity}.
\newblock {\em JHEP}, 09:150, 2021.

\bibitem{Zhang:2021cea}
Hongsheng Zhang and Yang Huang.
\newblock {Spherical gravitational waves and quasi-spherical waves scattered
  from black string in massive gravity}.
\newblock {\em JHEP}, 12:056, 2021.

\bibitem{Zhang:2021sjx}
Hongsheng Zhang.
\newblock {Non perturbative spherical gravitational waves}.
\newblock {\em Phys. Lett. B}, 816:136220, 2021.

\bibitem{Zhang:2014ala}
Hongsheng Zhang and Xin-Zhou Li.
\newblock {From thermodynamics to the solutions in gravity theory}.
\newblock {\em Phys. Lett. B}, 737:395--400, 2014.

\bibitem{Zhang:2014kla}
Hongsheng Zhang, Dao-Jun Liu, and Xin-Zhou Li.
\newblock {Black holes and gravitational waves in three-dimensional f(R)
  gravity}.
\newblock {\em Phys. Rev. D}, 90:124051, 2014.

\bibitem{Lu:2015cqa}
H.~Lu, A.~Perkins, C.~N. Pope, and K.~S. Stelle.
\newblock {Black Holes in Higher-Derivative Gravity}.
\newblock {\em Phys. Rev. Lett.}, 114(17):171601, 2015.

\bibitem{PhysRevD.97.024015}
Kevin Goldstein and James~Junior Mashiyane.
\newblock Ineffective higher derivative black hole hair.
\newblock {\em Phys. Rev. D}, 97:024015, Jan 2018.

\bibitem{Cai:2015fia}
Yi-Fu Cai, Gong Cheng, Junyu Liu, Min Wang, and Hezi Zhang.
\newblock {Features and stability analysis of non-Schwarzschild black hole in
  quadratic gravity}.
\newblock {\em JHEP}, 01:108, 2016.

\bibitem{Held:2022abx}
Aaron Held and Jun Zhang.
\newblock {Instability of spherically-symmetric black holes in Quadratic
  Gravity}.
\newblock 9 2022.

\bibitem{Novikov1965}
I.~D. Novikov.
\newblock Higher-derivative gravity, surface terms, and string theory.
\newblock {\em Soviet Ast.}, 8:857--863, 1965.

\bibitem{Podolsky:2019gro}
Jiri Podolsk\'y, Robert \v{S}varc, Vojtech Pravda, and Alena Pravdova.
\newblock {Black holes and other exact spherical solutions in Quadratic
  Gravity}.
\newblock {\em Phys. Rev. D}, 101(2):024027, 2020.

\bibitem{Pravda:2020zno}
Vojtech Pravda, Alena Pravdova, Jiri Podolsky, and Robert Svarc.
\newblock {Black holes and other spherical solutions in quadratic gravity with
  a cosmological constant}.
\newblock {\em Phys. Rev. D}, 103(6):064049, 2021.

\bibitem{Svarc:2018coe}
Robert Svarc, Jiri Podolsky, Vojtech Pravda, and Alena Pravdova.
\newblock {Exact black holes in quadratic gravity with any cosmological
  constant}.
\newblock {\em Phys. Rev. Lett.}, 121(23):231104, 2018.

\bibitem{Lin:2016jjl}
Kai Lin, A.~B. Pavan, G.~Flores-Hidalgo, and E.~Abdalla.
\newblock {New Electrically Charged Black Hole in Higher Derivative Gravity}.
\newblock {\em Braz. J. Phys.}, 47(4):419--425, 2017.

\bibitem{Wu:2019uvq}
Chao Wu, De-Cheng Zou, and Ming Zhang.
\newblock {Charged black holes in the Einstein-Maxwell-Weyl gravity}.
\newblock {\em Nucl. Phys. B}, 952:114942, 2020.

\bibitem{Zou:2020msk}
De-Cheng Zou, Chao Wu, Ming Zhang, and Ruihong Yue.
\newblock {Quasinormal modes of charged black holes in Einstein-Maxwell-Weyl
  gravity}.
\newblock {\em Chin. Phys. C}, 44(5):055102, 2020.

\bibitem{PhysRevD.102.124026}
S.~N. Sajadi, Robert~B. Mann, N.~Riazi, and Saeed Fakhry.
\newblock Analytically approximate solutions to higher derivative gravity.
\newblock {\em Phys. Rev. D}, 102:124026, Dec 2020.

\bibitem{Sajadi:2022tgi}
Seyed~Naseh Sajadi, Ali Hajilou, and Seyed~Hossein Hendi.
\newblock {Analytically approximation solution to $R^{2}$ gravity}.
\newblock {\em Eur. Phys. J. C}, 83(1):45, 2023.

\bibitem{PhysRevD.82.104026}
William Nelson.
\newblock Static solutions for fourth order gravity.
\newblock {\em Phys. Rev. D}, 82:104026, Nov 2010.

\bibitem{Fan:2014ala}
Zhong-Ying Fan and H.~Lu.
\newblock {Thermodynamical First Laws of Black Holes in Quadratically-Extended
  Gravities}.
\newblock {\em Phys. Rev. D}, 91(6):064009, 2015.

\bibitem{Hawking:1974rv}
S.~W. Hawking.
\newblock {Black hole explosions}.
\newblock {\em Nature}, 248:30--31, 1974.

\bibitem{AHARONOV198751}
Y.~Aharonov, A.~Casher, and S.~Nussinov.
\newblock The unitarity puzzle and planck mass stable particles.
\newblock {\em Physics Letters B}, 191(1):51--55, 1987.

\bibitem{Giddings:1992hh}
Steven~B. Giddings.
\newblock {Black holes and massive remnants}.
\newblock {\em Phys. Rev. D}, 46:1347--1352, 1992.

\bibitem{Banks:1992is}
Tom Banks, M.~O'Loughlin, and Andrew Strominger.
\newblock {Black hole remnants and the information puzzle}.
\newblock {\em Phys. Rev. D}, 47:4476--4482, 1993.

\bibitem{Xiang:2006mg}
Li~Xiang.
\newblock {A Note on the black hole remnant}.
\newblock {\em Phys. Lett. B}, 647:207--210, 2007.

\bibitem{Maziashvili:2005pp}
Michael Maziashvili.
\newblock {Black hole remnants due to GUP or quantum gravity?}
\newblock {\em Phys. Lett. B}, 635:232--234, 2006.

\bibitem{Chen:2012tx}
Deyou Chen and Xiaoxiong Zeng.
\newblock {The Schwarzschild black hole's remnant via the Bohr-Sommerfeld
  quantization rule}.
\newblock {\em Gen. Rel. Grav.}, 45:631--641, 2013.

\bibitem{Horowitz:2003he}
Gary~T. Horowitz and Juan~Martin Maldacena.
\newblock {The Black hole final state}.
\newblock {\em JHEP}, 02:008, 2004.

\bibitem{Xiang:2013sza}
Li~Xiang, Yi~Ling, and You~Gen Shen.
\newblock {Singularities and the Finale of Black Hole Evaporation}.
\newblock {\em Int. J. Mod. Phys. D}, 22:1342016, 2013.

\bibitem{Ali:2014xqa}
Ahmed~Farag Ali.
\newblock {Black hole remnant from gravity\textquoteright{}s rainbow}.
\newblock {\em Phys. Rev. D}, 89(10):104040, 2014.

\bibitem{Olmo:2013gqa}
Gonzalo~J. Olmo, D.~Rubiera-Garcia, and Helios Sanchis-Alepuz.
\newblock {Geonic black holes and remnants in Eddington-inspired Born-Infeld
  gravity}.
\newblock {\em Eur. Phys. J. C}, 74:2804, 2014.

\bibitem{Liu:2019lul}
Jing Liu, Zong-Kuan Guo, and Rong-Gen Cai.
\newblock {Primordial Black Holes from Cosmic Domain Walls}.
\newblock {\em Phys. Rev. D}, 101(2):023513, 2020.

\bibitem{Liu:2021svg}
Jing Liu, Ligong Bian, Rong-Gen Cai, Zong-Kuan Guo, and Shao-Jiang Wang.
\newblock {Primordial black hole production during first-order phase
  transitions}.
\newblock {\em Phys. Rev. D}, 105(2):L021303, 2022.

\bibitem{Hashino:2022tcs}
Katsuya Hashino, Shinya Kanemura, Tomo Takahashi, and Masanori Tanaka.
\newblock {Probing first-order electroweak phase transition via primordial
  black holes in the effective field theory}.
\newblock 11 2022.

\bibitem{Hawking:1971ei}
Stephen Hawking.
\newblock {Gravitationally collapsed objects of very low mass}.
\newblock {\em Mon. Not. Roy. Astron. Soc.}, 152:75, 1971.

\bibitem{Carr:1974nx}
Bernard~J. Carr and S.~W. Hawking.
\newblock {Black holes in the early Universe}.
\newblock {\em Mon. Not. Roy. Astron. Soc.}, 168:399--415, 1974.

\bibitem{PhysRevD.101.023010}
Alexandre Arbey, J\'er\'emy Auffinger, and Joseph Silk.
\newblock Constraining primordial black hole masses with the isotropic gamma
  ray background.
\newblock {\em Phys. Rev. D}, 101:023010, Jan 2020.

\end{thebibliography}

	\appendix
	\section{Equations of $A(r)$ and $B(r)$} \label{Appedix: eqsAB}
	Here, we present the equations of functions $\left\lbrace A,B\right\rbrace $.
	\begin{equation}\label{Eq: Eq1}
	0=A_{,rr}+\left(\frac{2}{r}+A_{,r}+\frac{\Omega}{2}\right)\left(A_{,r}+B_{,r}\right)+\Omega A_{,r}+\frac{1}{r^2}\left(1-\frac{e^{-2B}}{N}\right)+\frac{1}{2} \left(\Omega_{,r}+\Omega^2\right)+\frac{2 \Omega}{r},
	\end{equation}
	\begin{equation}\label{Eq: Eq2}
	\begin{aligned}
	0=&\left(A_{,r}+\frac{\Omega}{2}-\frac{1}{r}\right)B_{,rr} +B_{,r}\left[3 \Omega A_{,r}+A_{,r}^2+\frac{A_{,r}}{r}+\frac{1}{r^2}\left(1-\frac{e^{-2B}}{N}\right)+\frac{5\Omega^2}{4}\right]\\
	&+B_{,r}^2 \left(2 A_{,r}+\Omega-\frac{1}{2 r}\right)
	+A_{,r}\left(\frac{2 \Omega}{r}+\frac{\Omega^2}{4}+\frac{\Omega_{,r}}{2}-\frac{e^{-2 B}}{N}\right)+\left(\frac{3}{2 r}-\Omega\right) A_{,r}^2\\
	&-A_{,r}^3+\Omega\left[\frac{1}{2 r^2}+\frac{\Omega_{,r}}{4}-\frac{e^{-2 B}}{2N}\left(\frac{1}{r^2}+1\right)\right]-\frac{e^{-2B}}{r N}\left(\frac{1}{r^2}+\frac{1}{2}\right) +\frac{e^{-4B}}{2rN^2}\\
	&+\frac{1}{r^3}-\frac{\Omega_{,r}}{2 r}+\Omega^2\left(\frac{1}{2r}+\frac{\Omega}{4}\right),
	\end{aligned}
	\end{equation}
	where $\Omega(r)=N_{,r}/N$.

\end{document}